\documentclass[11pt,a4paper]{article}

\usepackage[margin=1in]{geometry}
\usepackage{amsmath,amssymb,mathtools,bm}
\usepackage{mathrsfs}
\usepackage{physics}
\usepackage{hyperref}
\hypersetup{hidelinks}
\usepackage{enumitem}
\usepackage{microtype}
\usepackage{booktabs}
\usepackage{hyperref}
\usepackage{amsmath}
\usepackage{hyperref}
\usepackage{comment}
\usepackage{enumitem}
\usepackage{listings}
\usepackage[section]{placeins} 
\usepackage[version=4]{mhchem}
\usepackage{comment}
\usepackage{textcomp}
\usepackage{gensymb}
\usepackage{caption}
\usepackage{dsfont}
\usepackage{amssymb}
\usepackage{subcaption}
\usepackage{tikz-cd} 
\usepackage{subcaption}
\usepackage{accents}
\usepackage{svg}
\usepackage[document]{ragged2e}
\usepackage{acronym}
\usepackage{tikz-feynman}
\usepackage{feynmp-auto}
\usepackage{afterpage}
\usepackage{blkarray}
\usepackage{booktabs}
\usepackage{slashed}
 
\usepackage[utf8]{inputenc}
\usepackage{comment}
\usepackage[T1]{fontenc}
\usepackage{lmodern}   
\usepackage{mathrsfs}
\usepackage{authblk}
\numberwithin{equation}{section}

\usepackage[dvipsnames]{xcolor}

\newcommand{\Lag}{\mathcal{L}}
\newcommand{\Kphi}{\mathcal{K}_{\phi}}
\newcommand{\DA}{\mathcal{D}_{A}}
\newcommand{\Delphi}{\Delta_{\phi}}
\newcommand{\Ahat}{\widehat A}
\newcommand{\Fhat}{\widehat F}
\newcommand{\phihat}{\widehat\phi}
\newcommand{\Dhat}{\widehat D}
\newcommand{\eps}{\varepsilon}
\newcommand{\cJ}{\mathcal{J}}
\newcommand{\cV}{\mathcal{V}}
\newcommand{\ii}{\mathrm{i}}
\newcommand{\PT}{P_{\rm T}}
\newcommand{\PL}{P_{\rm L}}
\newcommand{\phibarhat}{\widehat\phi^{\dagger}}

\title{\textbf{Berends-Giele currents for higher derivative gauge theories}}
\author[a]{Pratik Chattopadhyay\thanks{pratikpc@gmail.com}}
\affil[a]{\textit{Department of Physics, Ariel University, Ariel 4070000, Israel}}

\begin{document}
\maketitle

\begin{abstract}
We derive Berends-Giele (BG) currents for higher derivative scalar QED (HDQED) and Yang-Mills (HDYM). In HDQED, we first compute the propagators and the vertices from the higher derivative Lagrangian and discuss the pole structures of the photon and the scalar. We then compute the BG currents using the Feynman diagram setup and prove the Ward identities for these currents at two and three points. Next, we used the perturbiner formalism to re-compute these currents and show that they are consistent with the Feynman diagram approach. In a similar fashion, we derive the BG currents in higher derivative Yang-Mills and find that remarkably, these currents are equal to pure Yang-Mills currents when all external on-shell legs are massless gluons. We prove this in all generality using the BG recursion and explicitly show this at lower points. However, the genuine higher derivative structure enters when one or many on-shell states are the so called massive Merlin modes. Using the multi-particle currents, we reproduce the result that all tree level amplitudes with one Merlin particle vanishes and the three and four Merlin amplitudes match the previously computed result. The future prospects of BG recursion for higher derivative gravity and off-shell colour- kinematics duality is also discussed. 

\end{abstract}
\section{Introduction} 
Gauge theories, both abelian and non-abelian are cornerstones of particle physics. They have shaped how we think of interactions of fundamental particles and address novel applications in condensed matter systems. On the other hand, general relativity explains gravitational interactions predominantly at cosmological scales. Both these theories come with issues of ultraviolet divergences when loop corrections are taken into account. While gauge theories like QED and QCD are renormalizable, QED admits the Landau pole problem in the UV while QCD becomes strongly coupled at the infrared, inviting for better methods for computations. Gravity on the other hand is non-renormalizable and certainly cries for modifications at the ultraviolet regime. These issues motivated people to construct higher derivative field theories, which do not just modify the interactions, but change the spectrum of the theory. Thus, if we consider higher derivative scalar QED (HDQED) or Yang-Mills (HDYM), in addition to massless poles of the gauge sector, there are massive poles introduced in these kinds of theories. Lee Wick quantum field theories are certainly the most striking examples of such a class of theories, see \cite{LeeWick1970}-\cite{Neves}. The massive photon in HDQED has
a problem which is that its residue has the wrong sign, which leads to violating causality. Nonetheless, it can be shown that the unitarity of the S-matrix can be preserved via
deformation of the integration contour as long as the massive photon decays quickly thereby making sense of the theory. Similar considerations hold for HDYM. Inspite of such behaviors, these theories, in particular HDYM is interesting because it is related to quadratic gravity at the level of scattering amplitudes. While scattering amplitudes have a long history, they have been flooding since the last two decades owing to several interesting relations and mathematical structures which were and are being discovered. One of these relations is the famous BCJ double copy \cite{Bern} that was initially observed for pure Yang-Mills and gravity and later generalised to many more classes of theories, and more recently in higher derivative QCD and quadratic gravity, see \cite{Menezes}. The double copy is a strikingly beautiful result which holds at the on-shell level. However, its off-shell manifestations are yet to be discovered in their full glory. An interesting off-shell object of field theories is the Berends-Giele (BG) current. This current satisfies the so called Berends-Giele recursion formula so that higher point currents can be obtained from lower points, making the task of computation in control. These currents can also be used to compute tree level amplitudes and even loop level integrands. Thus, they are a contemporary tool in modern amplitude methods, see \cite{pratik}-\cite{Lopez}. The recent discovery of the double copy at the level of these currents \cite{Joris} motivates us to analyze the structure of these currents in HDQED and HDYM. As illustrated in \cite{Menezes}, HDYM double copies to quadratic gravity at the amplitude level. However, the off-shell generalization of this double copy is not yet understood in such theories. In this work, we thus begin with a modest aim to analyze the BG currents in HDQED and HDYM with particular emphasis on the interplay between their massless and massive propagator branches. As we shall see, the purely massless sector exhibits striking cancellations of the higher-derivative poles, whereas currents involving massive states retain genuinely new higher-derivative structures. In \cite{Menezes}, it was shown that tree amplitudes in HDYM with all massless external gluon legs are the same as in pure YM and that with one massive external Merlin leg vanishes. We give an off-shell manifestation of the same idea at the level of BG currents. Thus, we show and prove that even the BG currents with all external massless on-shell legs in HDYM are the same as pure YM. For one Merlin massive on-shell leg, the amplitude vanishes as a result of the structure of the BG current. Since these currents are the building blocks of tree level amplitudes and follow a recursion relation, our proof is a more fundamental explanation of why such a phenomenon happens in HDYM. There has been a lot of work in the higher derivative and non-local sector of YM, with particular emphasis on leading singularities and tree level amplitudes, see \cite{Boos}-\cite{Zhang}. However, the loop level amplitudes of these sectors are still at a nascent stage. BG currents are a helpful tool to build loop level integrands by sewing two off-shell legs in appropriate ways and therefore this present work can lead to understanding loop level recursion relations in higher derivative gauge theories and can also hint at understanding colour-kinematics duality both at the level of BG currents and loop level integrands. 
\\~\\
The rest of the paper is as follows. In section 2, we review Berends-Giele current briefly. In section 3 we write down the Lagrangian for HDQED and explain the Feynman rules. In the rest of the subsections we compute the BG currents in this theory using the diagrammatic approach. Then we prove the Ward identity for the 3 point current. We also write down the perturbiner expansion of the higher derivative theory and in the next subsequent subsection, we re-compute the lower point scalar and photon currents and show that they match the ones from Feynman diagrams. In section 4, and upto subsection 4.3, we write down the Lagrangian for HDYM and expand on the Feynman rules. Then in subsections 4.4 and 4.5, we compute the BG currents in this theory. In the next subsection, we give the perturbiner expansion and reproduce the low rank currents. Subsection 4.7 is one of the main results of this work where we prove the equality of the BG currents of YM and HDYM for all external massless gluon legs at arbitrary multiplicity. In the subsequent subsections, we compute the amplitudes of the currents and show that they match \cite{Menezes}. We conclude with a discussion in section 5.
\section{Review of Berends-Giele currents}
\label{sec:BG_review}

Berends-Giele recursion provides an efficient way of organizing tree-level
scattering amplitudes in gauge theories in terms of multiparticle off-shell
currents.  Originally introduced for Yang-Mills theory, the method replaces a
direct summation over a rapidly growing number of Feynman diagrams by a
recursive construction in which lower-multiplicity currents are combined to
form higher-multiplicity ones.  In this sense, Berends-Giele currents provide
an intermediate description between ordinary off-shell Green functions and
fully on-shell scattering amplitudes.  They retain one distinguished off-shell
leg while the remaining external states are kept on shell.

Consider a collection of color-ordered external gluons labelled by an ordered
word
\begin{equation}
P=p_1p_2\cdots p_n .
\end{equation}
Associated with this word is a total momentum
\begin{equation}
K_P=\sum_{i\in P}k_i,
\end{equation}
and a multiparticle current $J_P^\mu$.  The current represents the sum of all
tree diagrams containing the ordered on-shell external states belonging to
$P$ and one additional off-shell leg carrying momentum $K_P$.  The off-shell
leg is not interpreted as an external physical particle; rather, it provides
the line through which the current can be attached to another vertex in a
higher-point diagram.  At rank one the current is simply the polarization
vector of an external gluon,
\begin{equation}
J_i^\mu=\varepsilon_i^\mu.
\end{equation}

The essential idea of the recursion is that a higher-rank current may be
obtained by partitioning its ordered set of external legs into smaller
ordered subsets and connecting the corresponding lower-rank currents through
the interaction vertices of the theory.  In ordinary Yang--Mills theory only
cubic and quartic elementary gauge-field vertices are present.  Therefore a
multiparticle current receives contributions from two types of ordered
decompositions,
\begin{equation}
P=QR,
\qquad
P=QRS,
\end{equation}
corresponding respectively to cubic and quartic interactions.  Schematically,
the Yang--Mills recursion may be represented as
\begin{equation}
J_P
\sim
\frac{1}{K_P^2}
\left[
\sum_{P=QR}V_3(J_Q,J_R)
+
\sum_{P=QRS}V_4(J_Q,J_R,J_S)
\right].
\label{eq:BG_YM_schematic}
\end{equation}
The factor $1/K_P^2$ is the propagator associated with the off-shell root of
the current.  The objects $J_Q$, $J_R$, and $J_S$ already contain all tree
subdiagrams associated with the corresponding lower-rank ordered words.
Thus the recursive construction automatically generates the complete set of
tree diagrams without having to enumerate them individually.

For example, the rank-two current $J_{12}^\mu$ contains a single cubic
interaction joining the external states $1$ and $2$ to an off-shell leg.  At
rank three, $J_{123}^\mu$ receives two exchange contributions, corresponding
to the ordered channels $(12)|3$ and $1|(23)$, together with the quartic
contact interaction.  At higher multiplicity the same pattern continues, i.e 
the recursion repeatedly joins previously constructed currents through the
elementary vertices of the theory.  This hierarchical organization is one of
the principal advantages of the Berends-Giele method.

The off-shell nature of the currents is important.  Individual
Berends-Giele currents are gauge dependent and are not themselves physical
observables.  Nevertheless, they contain the complete information required
to reconstruct tree-level scattering amplitudes.  Once a current involving
$(n-1)$ external particles has been constructed, the corresponding $n$-point
amplitude is obtained by amputating the propagator associated with its
off-shell leg, putting this off-shell momentum on shell, and contracting the
remaining Lorentz index with the polarization vector of the final external
state.  In this way the gauge dependence of the intermediate currents drops
out of the physical amplitude.

Berends-Giele recursion is also closely related to the perturbiner
description of classical field equations.  In the perturbiner approach the
classical gauge field is expanded as a formal multiparticle plane-wave
series,
\begin{equation}
A^\mu(x)
=
\sum_{P\neq\varnothing}
J_P^\mu\,T^P\,e^{iK_P\cdot x},
\label{eq:BG_perturbiner}
\end{equation}
where $T^P$ denotes an ordered product of color generators.  Substitution of
this expansion into the classical equations of motion and comparison of the
coefficients multiplying each word $P$ produces recursion relations for the
multiparticle coefficients $J_P^\mu$.  These coefficients are precisely the
Berends-Giele currents.  The perturbiner construction therefore provides an
alternative derivation of the same recursive structure that follows
from the diagrammatic approach.

This equivalence between the Feynman-diagram and perturbiner viewpoints is
particularly useful when the theory contains non-standard kinetic operators
or higher-derivative interactions. In the diagrammatic language, such
modifications change both the propagators and the interaction vertices, so
the recursive structure may involve additional poles and higher-point
contact terms.  In the perturbiner formulation the same information is
encoded directly in the nonlinear field equations, and all interaction
orders generated by the action are incorporated automatically. The two
descriptions therefore provide complementary checks on the resulting
multi-particle currents.

Another useful feature of the Berends-Giele framework is that it makes the
analytic structure of amplitudes visible already at the current level.
Poles associated with intermediate propagating states appear explicitly in
the multi-particle currents, while contact interactions are separated from
factorization channels.  Consequently, if a theory contains more than one
propagating branch, the corresponding currents can be decomposed according
to the different poles before the final on-shell amplitude is taken.  This
observation will play an important role in the higher-derivative gauge theories considered below.

\section{Higher-derivative scalar QED}

The action of higher derivative scalar QED, often called Lee-Wick scalar QED is 
\begin{equation}
\Lag_{\rm HD}
=
-\frac14\Fhat_{\mu\nu}\Fhat^{\mu\nu}
+\frac{1}{2M_A^2}
(\partial^\mu\Fhat_{\mu\nu})(\partial_\rho\Fhat^{\rho\nu})
+(\Dhat_\mu\phihat)^*\Dhat^\mu\phihat
-\frac1{M_\phi^2}(\Dhat^2\phihat)^*(\Dhat^2\phihat)
-m^2\phihat^*\phihat .
\label{eq:HDLag}
\end{equation}
where $M_A$ and $M_\phi$ are the higher derivative scales of the photon and the scalar respectively. It is clear from the above expression that the second term in the Lagrangian is the higher derivative term for the photon and the fourth term is the higher derivative term for the scalar. To clarify notations, the covariant derivative and the field strength are given by
\begin{equation}
\Dhat_\mu=\partial_\mu-\ii g\Ahat_\mu,
\qquad
\Fhat_{\mu\nu}=\partial_\mu\Ahat_\nu-\partial_\nu\Ahat_\mu .
\end{equation}
Because the theory is Abelian, $\partial^\mu\Fhat_{\mu\nu}$ is linear in
$\Ahat_\mu$.  Hence the higher-derivative photon sector is still quadratic and
all interactions arise from the charged scalar sector.
For inversion of the gauge quadratic operator we add the standard covariant
gauge-fixing term
\begin{equation}
\Lag_{\rm gf}=-\frac{1}{2\xi}(\partial\cdot\Ahat)^2.
\label{eq:gaugeFixing}
\end{equation}
\subsection{Higher-derivative propagators}

\subsubsection{Scalar propagator}
In momentum space, the scalar propagator is
\begin{equation}
\Delphi(p)=\frac{1}{p^2-\frac{p^4}{M_\phi^2}-m^2}.
\label{eq:scalarProp}
\end{equation}
The two scalar poles are the roots of $\Kphi(s)=0$,
\begin{equation}
m_\pm^2
=
\frac{M_\phi^2}{2}
\left(1\pm\sqrt{1-\frac{4m^2}{M_\phi^2}}\right).
\end{equation}
\subsubsection{Photon propagator}
\label{sec:exactPhotonProp}

We use the higher-derivative photon propagator in exactly the form given in (6) of \cite{AbuAjamiehLWQED}:
\begin{equation}
\Pi_g^{\mu\nu}(k)
=
\frac{\ii M_A^2}{k^2(k^2-M_A^2)}
\left[
\eta^{\mu\nu}
-(1-\xi)\frac{k^\mu k^\nu}{k^2}
-\xi\frac{k^\mu k^\nu}{M_A^2}
\right].
\label{eq:paperPhotonProp}
\end{equation}
Let us introduce the transverse and longitudinal projectors
\begin{equation}
P_T^{\mu\nu}(k)
\equiv
\eta^{\mu\nu}-\frac{k^\mu k^\nu}{k^2},
\qquad
P_L^{\mu\nu}(k)
\equiv
\frac{k^\mu k^\nu}{k^2},
\label{eq:PTPL}
\end{equation}
Using these projectors, the useful form of the propagator is
\begin{equation}
\Pi_g^{\mu\nu}(k)
=-\ii\left[
\frac{P_T^{\mu\nu}(k)}{\DA(k^2)}
+\frac{\xi P_L^{\mu\nu}(k)}{k^2}
\right],
\qquad
\DA(s)\equiv s\left(1-\frac{s}{M_A^2}\right).
\label{eq:projectorPhotonProp}
\end{equation}
Thus the Lee--Wick factor modifies only the transverse sector.  The
longitudinal gauge mode retains the ordinary covariant-gauge denominator
$k^2$.
In the BG analysis below we suppress the common overall propagator
factor $-\ii$, and so it is convenient to define the stripped tensor propagator
\begin{equation}
\mathscr G_A^{\mu\nu}(k)
\equiv
\frac{P_T^{\mu\nu}(k)}{\DA(k^2)}
+\frac{\xi P_L^{\mu\nu}(k)}{k^2}.
\label{eq:strippedPhotonProp}
\end{equation}
The two physical transverse poles of the same higher-derivative field are
\begin{equation}
k^2=0,
\qquad
k^2=M_A^2.
\end{equation}

\subsection{Higher-derivative scalar--photon vertices}
\label{sec:vertices}
\subsubsection{$\phihat\phihat\Ahat$ vertex}
Let a scalar of momentum $p$ absorb a photon of momentum $q$ and emerge with momentum $p'$.
Let us define a function $C(a,b)=1-\frac{a^2+b^2}{M_{\phi}^2}$.
The resulting cubic vertex (upto numerical factors) is
\begin{equation}
\cV_1^\mu(p',p)
=(p'+p)^\mu\,C(p'^2,p^2)
=(p'+p)^\mu
\left[1-\frac{p'^2+p^2}{M_\phi^2}\right].
\label{eq:V1}
\end{equation}
The corresponding Feynman rule is $\ii g\cV_1^\mu$ in our convention.
\subsubsection{$\phihat\phihat\Ahat\Ahat$ vertex vertex}
Let two photons with momenta $q_2,q_3$ enter the scalar line of momentum $p$, with another scalar emerging with momentum $p'$. Then the symmetric stripped vertex is
\begin{equation}
\begin{aligned}
\cV_2^{\mu\nu}(p',p;q_2,q_3)
={}&
2\eta^{\mu\nu}C(p'^2,p^2)
\\
&-\frac1{M_\phi^2}
\Big[
(2p+q_2)^\mu(2p+2q_2+q_3)^\nu
\\
&\hspace{23mm}
+(2p+2q_3+q_2)^\mu(2p+q_3)^\nu
\Big].
\end{aligned}
\label{eq:V2}
\end{equation}
To use this Feynman rule, we need to multiply an additional factor of $ig^2$. This vertex is symmetric under
$(q_2,\mu)\leftrightarrow(q_3,\nu)$.
\subsubsection{Three and four photon scalar vertices}

For the low-rank currents below we only need $\cV_1$ and $\cV_2$. Nevertheless, the all-rank recursion also contains the three- and four-photon
vertices.  Rather than writing unnecessarily long momentum-space tensors, we
define them by using the equations \eqref{higher vertices} in Appendix A.
\begin{equation}
\begin{aligned}
\ii g^3\cV_3\equiv \text{the cubic-in-}\Ahat\text{ Feynman rule generated by }\mathcal K_3,\nonumber\\
\ii g^4\cV_4\equiv \text{the quartic-in-}\Ahat\text{ Feynman rule generated by }\mathcal K_4.
\end{aligned}
\end{equation}

\subsection{BG-current conventions}

We first compute the low-rank currents directly from tree Feynman diagrams.
The following convention is convenient.  For a scalar seed labelled $1$ and
a set $P$ of external photon labels, we define
$\Phi_{1|P}$
to be the amputated external on-shell states joined to an off-shell scalar line,
including the final off-shell scalar propagator.  For a scalar-antiscalar
pair $1,\bar2$ and additional photon labels $P$, we define
$\Ahat_{1\bar2|P}^\mu$
to be the corresponding off-shell photon current, including the full
higher-derivative photon propagator.
For compactness let us also define
\begin{equation}
D_P\equiv \Kphi(k_P^2),
\qquad
\Delta_P\equiv\frac1{D_P},
\qquad
\Pi_P\equiv\frac1{\DA(k_P^2)}.
\label{eq:Ddefs}
\end{equation}

\subsection{Scalar BG currents from Feynman diagrams}
\label{sec:scalarDiagrams}

\subsubsection{Two-particle scalar current}

We take an on-shell scalar with momentum$k_1$ and one photon of momentum$k_2$.  The off-shell scalar
momentum is
\begin{equation}
k_{12}=k_1+k_2.
\end{equation}
There is a single graph.  With the stripped-current convention inherited from
$\ii\Delta\times\ii g\cV_1$, we define the BG coefficient as,
\begin{equation}
\Phi_{1|2}
=
\frac{g\,\cV_1^\mu(k_{12},k_1)\eps_{2\mu}}{D_{12}}.
\label{eq:Phi12Diagram}
\end{equation}
Using \eqref{eq:V1},
\begin{equation}
\Phi_{1|2}
=
\frac{g\,(2k_1+k_2)\cdot\eps_2}{D_{12}}
C(k_{12}^2,k_1^2).
\label{eq:Phi12Explicit}
\end{equation}
For a transverse elementary photon seed,
$k_2\cdot\eps_2=0$, this becomes
\begin{equation}
\Phi_{1|2}
=
\frac{2g\,k_1\cdot\eps_2}{D_{12}}
C(k_{12}^2,k_1^2).
\label{eq:Phi12Transverse}
\end{equation}
In the above, we do not decide whether the external photon is on $k_2^2=0$ or $k_2^2=M_A^2$. In either case, it returns the same BG current by virtue of the transverse property of the photon attached.

\subsubsection{Three-particle scalar current}

Now we take a scalar with momentum $k_1$ and two photons with momenta $k_2,k_3$. Let us define
\begin{equation}
k_{12}=k_1+k_2,
\qquad
k_{13}=k_1+k_3,
\qquad
k_{123}=k_1+k_2+k_3.
\end{equation}
We note that there are exactly three tree diagrams. The first diagram has the scalar and a photon with momentum $k_3$ entering into a vertex, with an outgoing scalar and the photon with $k_2$ inserted on this last scalar line.
The intermediate scalar carries $k_{13}$.  The contribution is
\begin{align}
\Phi_{1|23}^{(a)}
={}&
\Delta_{123}
\big[g\cV_1^\mu(k_{123},k_{13})\eps_{2\mu}\big]
\Delta_{13}
\big[g\cV_1^\nu(k_{13},k_1)\eps_{3\nu}\big]
\\
={}&
\frac{g^2}{D_{123}D_{13}}
\big[(k_{123}+k_{13})\cdot\eps_2\big]
\big[(2k_1+k_3)\cdot\eps_3\big]
\nonumber\\
&\hspace{13mm}\times
C(k_{123}^2,k_{13}^2)
C(k_{13}^2,k_1^2).
\label{eq:Phi123a}
\end{align}
The second diagram is where the photons with momenta $k_2$ and $k_3$ are exchanged. This gives
\begin{align}
\Phi_{1|23}^{(b)}
={}&
\frac{g^2}{D_{123}D_{12}}
\big[(k_{123}+k_{12})\cdot\eps_3\big]
\big[(2k_1+k_2)\cdot\eps_2\big]
\nonumber\\
&\hspace{13mm}\times
C(k_{123}^2,k_{12}^2)
C(k_{12}^2,k_1^2).
\label{eq:Phi123b}
\end{align}
The last diagram is the contact graph which contains the two photon-two scalar vertex, namely, $\ii g^2\cV_2$ .  Relative to the two sequential exchange graphs it
enters with the opposite sign. This then gives
\begin{equation}
\Phi_{1|23}^{(c)}
=-\frac{g^2}{D_{123}}
\cV_2^{\mu\nu}(k_{123},k_1;k_2,k_3)
\eps_{2\mu}\eps_{3\nu}.
\label{eq:Phi123c}
\end{equation}
Using \eqref{eq:V2},
\begin{align}
\cV_2^{\mu\nu}\eps_{2\mu}\eps_{3\nu}
={}&
2(\eps_2\cdot\eps_3)C(k_{123}^2,k_1^2)
\nonumber\\
&-\frac1{M_\phi^2}
\Big[
(2k_1+k_2)\cdot\eps_2\,
(2k_1+2k_2+k_3)\cdot\eps_3
\nonumber\\
&\hspace{22mm}
+(2k_1+2k_3+k_2)\cdot\eps_2\,
(2k_1+k_3)\cdot\eps_3
\Big].
\label{eq:V2contracted}
\end{align}

The complete three-particle scalar current is obtained by adding the three diagrams above. This gives
\begin{equation}
\begin{aligned}
\Phi_{1|23}
=\frac{g^2}{D_{123}}
\Bigg\{&
\frac{[(k_{123}+k_{13})\cdot\eps_2][(2k_1+k_3)\cdot\eps_3]}{D_{13}}
C(k_{123}^2,k_{13}^2)C(k_{13}^2,k_1^2)
\\
&+
\frac{[(k_{123}+k_{12})\cdot\eps_3][(2k_1+k_2)\cdot\eps_2]}{D_{12}}
C(k_{123}^2,k_{12}^2)C(k_{12}^2,k_1^2)
\\
&-2(\eps_2\cdot\eps_3)C(k_{123}^2,k_1^2)
\\
&+\frac1{M_\phi^2}
\Big[
(2k_1+k_2)\cdot\eps_2\,
(2k_1+2k_2+k_3)\cdot\eps_3
\\
&\hspace{18mm}
+(2k_1+2k_3+k_2)\cdot\eps_2\,
(2k_1+k_3)\cdot\eps_3
\Big]
\Bigg\}.
\end{aligned}
\label{eq:Phi123Final}
\end{equation}
The result is manifestly symmetric under $2\leftrightarrow3$.

\subsection{Photon BG currents from Feynman diagrams}
\label{sec:photonDiagrams}

The Abelian gauge sector has no pure-photon interaction.  Therefore every
nontrivial composite photon current is sourced by charged scalar lines. 

\subsubsection{Two-particle photon BG current}

Let $k_1$ denote the momentum of an outgoing scalar and $k_2$ for an outgoing antiscalar. We insert the off shell photon in this scalar-scalar-photon vertex. The vertex is given by 
\begin{equation}
\cJ_{1\bar2}^\mu
=g\cV_1^\mu(k_1,-k_2)
=g(k_1-k_2)^\mu C(k_1^2,k_2^2).
\label{eq:J12source}
\end{equation}
Attaching the off shell propagator of the photon, we obtain the BG current as
\begin{equation}
\Ahat_{1\bar2}^\mu
=
\left[
\frac{P_T^{\mu\nu}(K)}{\DA(K^2)}
+\frac{\xi P_L^{\mu\nu}(K)}{K^2}
\right]
\cJ_{1\bar2,\nu}.
\label{eq:A12Projector}
\end{equation}

\subsubsection*{Ward identity for 2-current}

The one-photon Ward identity gives 
\begin{align}
K_\mu\cJ_{1\bar2}^\mu
&=gK_\mu\cV_1^\mu(k_1,-k_2)
\nonumber\\
&=g\left[\Kphi(k_1^2)-\Kphi(k_2^2)\right].
\label{wardid}
\end{align}
Since the scalar and antiscalar are on-shell, they satisfy
\begin{equation}
\Kphi(k_1^2)=\Kphi(k_2^2)=0,
\end{equation}
then
\begin{equation}
K_\mu\cJ_{1\bar2}^\mu=0.
\label{eq:rank2Ward}
\end{equation}
Hence the longitudinal part of Eq.~\eqref{eq:A12Projector} vanishes and
\begin{equation}
\Ahat_{1\bar2}^\mu
=
\frac{\cJ_{1\bar2}^\mu}{\DA(K^2)}
=
\frac{g(k_1-k_2)^\mu C(k_1^2,k_2^2)}
{K^2(1-K^2/M_A^2)}.
\label{eq:A12HD}
\end{equation}
Thus the simple scalar denominator used for the rank-two photon current is a consequence of the Ward identity. It is not the full propagator before contraction with the source.

\subsubsection{Three-particle photon source}
For the three particle current, we add the second photon in three possible ways to the cubic vertex. First, the convention is as follows. Let
\begin{equation}
K=k_{123}=k_1+k_2+k_3,
\qquad
q_0=-K.
\end{equation}
The scalar line has momentum
\begin{equation}
p=k_1,
\qquad
p'=-k_2,
\qquad
p'=p+q_0+k_3.
\label{eq:rank3routing}
\end{equation}
The two possible intermediate scalar momenta are
\begin{equation}
r_3=p+k_3=k_1+k_3=k_{13},
\qquad
r_0=p+q_0=-(k_2+k_3)=-k_{23}.
\end{equation}
The relevant scalar propagators are
\begin{equation}
\Delta_\phi(r_3)=\frac1{D_{13}},
\qquad
\Delta_\phi(r_0)=\frac1{D_{23}}.
\end{equation}
Given these, the three diagrams are as follows. In the first diagram, the photon with momentum $k_3$ is emitted from the scalar line with scalar momentum $k_1$. This results in a scalar propagator with momentum $k_{13}$ before final vertex and off-shell photon line is inserted into it. The diagram gives
\begin{equation}
\cJ_{(a)}^\mu
=g^2\eps_{3\nu}
\cV_1^\mu(p',r_3)\Delta_\phi(r_3)
\cV_1^\nu(r_3,p).
\label{eq:J123aCompact}
\end{equation}
Expanding the vertices gives
\begin{align}
\cJ_{(a)}^\mu
={}&
\frac{g^2}{D_{13}}
(k_1+k_3-k_2)^\mu
[(2k_1+k_3)\cdot\eps_3]
\nonumber\\
&\hspace{12mm}\times
C(k_2^2,k_{13}^2)C(k_{13}^2,k_1^2).
\label{eq:J123a}
\end{align}

In the second diagram, instead of the on-shell photon with momentum $k_3$, the off-shell photo is inserted into the scalar leg with scalar momentum $k_1$. This is the same diagram as before, but with the two photon legs being interchanged. This gives
\begin{equation}
\cJ_{(b)}^\mu
=g^2\eps_{3\nu}
\cV_1^\nu(p',r_0)\Delta_\phi(r_0)
\cV_1^\mu(r_0,p).
\label{eq:J123bCompact}
\end{equation}
In the all-outgoing momentum notation this is
\begin{align}
\cJ_{(b)}^\mu
={}&
-\frac{g^2}{D_{23}}
(k_1-k_2-k_3)^\mu
[(2k_2+k_3)\cdot\eps_3]
\nonumber\\
&\hspace{12mm}\times
C(k_2^2,k_{23}^2)C(k_{23}^2,k_1^2).
\label{eq:J123b}
\end{align}
The minus sign in \eqref{eq:J123b} is the consequence of rewriting
the oriented scalar momentum $r_0=-(k_2+k_3)$ in all-outgoing notation.
\\~\\
The third diagram is the contact diagram with quartic scalar-photon interaction. This gives
\begin{equation}
\cJ_{(c)}^\mu
=-g^2\eps_{3\nu}
\cV_2^{\mu\nu}(p',p;q_0,k_3).
\label{eq:J123c}
\end{equation}
Using \eqref{eq:V2}, one obtains
\begin{align}
\cJ_{(c)}^\mu
={}&
g^2\Bigg\{
-2\eps_3^\mu C(k_1^2,k_2^2)
\nonumber\\
&\quad+\frac1{M_\phi^2}
\Big[
-(k_1-k_2-k_3)^\mu(2k_2+k_3)\cdot\eps_3
\nonumber\\
&\hspace{29mm}
+(k_1-k_2+k_3)^\mu(2k_1+k_3)\cdot\eps_3
\Big]
\Bigg\}.
\label{eq:J123cExplicit}
\end{align}
Thus the complete 3 point photon BG current is given by the following expression
\begin{equation}
\begin{aligned}
\frac{\cJ_{1\bar2|3}^\mu}{g^2}
={}&
\frac{(k_1+k_3-k_2)^\mu[(2k_1+k_3)\cdot\eps_3]}{D_{13}}
C(k_2^2,k_{13}^2)C(k_{13}^2,k_1^2)
\\
&-
\frac{(k_1-k_2-k_3)^\mu[(2k_2+k_3)\cdot\eps_3]}{D_{23}}
C(k_2^2,k_{23}^2)C(k_{23}^2,k_1^2)
\\
&-2\eps_3^\mu C(k_1^2,k_2^2)
\\
&+\frac1{M_\phi^2}
\Big[
-(k_1-k_2-k_3)^\mu(2k_2+k_3)\cdot\eps_3
\\
&\hspace{17mm}
+(k_1-k_2+k_3)^\mu(2k_1+k_3)\cdot\eps_3
\Big].
\end{aligned}
\label{eq:J123Explicit}
\end{equation}

\subsection{Ward identity for 3-point current}
\label{sec:rank3Ward}

The cancellation is clearest if we use the compact oriented expressions
\eqref{eq:J123aCompact}--\eqref{eq:J123c}.  Since the root photon momentum
entering the scalar line is $q_0=-K$, it is enough to prove
$q_{0\mu}\cJ_{1\bar2|3}^\mu=0$.

For the first diagram, we use \eqref{wardid} on the first vertex as follows
\begin{align}
q_{0\mu}\cJ_{(a)}^\mu
={}&g^2\eps_{3\nu}
\left[\Kphi(p'^2)-\Kphi(r_3^2)\right]
\Delta_\phi(r_3)\cV_1^\nu(r_3,p)
\nonumber\\
={}&g^2\eps_{3\nu}
\left[
\frac{\Kphi(p'^2)}{\Kphi(r_3^2)}-1
\right]
\cV_1^\nu(r_3,p).
\label{eq:WardJa}
\end{align}
For an on-shell outgoing antiscalar, $\Kphi(p'^2)=\Kphi(k_2^2)=0$, hence
\begin{equation}
q_{0\mu}\cJ_{(a)}^\mu
=-g^2\eps_{3\nu}\cV_1^\nu(r_3,p).
\label{eq:WardJaOnShell}
\end{equation}

For the second diagram, the vertex attached to the on shell photon is $\cV_1^\mu(r_0,p)$. Here we use \eqref{wardid} to the vertex attached to the off-shell photon, thus giving
\begin{align}
q_{0\mu}\cJ_{(b)}^\mu
={}&g^2\eps_{3\nu}\cV_1^\nu(p',r_0)
\Delta_\phi(r_0)
\left[\Kphi(r_0^2)-\Kphi(p^2)\right].
\end{align}
Using the on-shell condition $\Kphi(p^2)=\Kphi(k_1^2)=0$ gives
\begin{equation}
q_{0\mu}\cJ_{(b)}^\mu
=g^2\eps_{3\nu}\cV_1^\nu(p',r_0).
\label{eq:WardJbOnShell}
\end{equation}

Finally the two-photon Ward identity, with $q_2=q_0$ and
$q_3=k_3$, reads
\begin{equation}
q_{0\mu}\cV_2^{\mu\nu}(p',p;q_0,k_3)
=
\cV_1^\nu(p',p+q_0)
-\cV_1^\nu(p'-q_0,p).
\end{equation}
Because
\begin{equation}
p+q_0=r_0,
\qquad
p'-q_0=r_3,
\end{equation}
we find
\begin{align}
q_{0\mu}\cJ_{(c)}^\mu
&=-g^2\eps_{3\nu}
\left[
\cV_1^\nu(p',r_0)-\cV_1^\nu(r_3,p)
\right]
\nonumber\\
&=-g^2\eps_{3\nu}\cV_1^\nu(p',r_0)
+g^2\eps_{3\nu}\cV_1^\nu(r_3,p).
\label{eq:WardJc}
\end{align}
Adding \eqref{eq:WardJaOnShell}, \eqref{eq:WardJbOnShell} and
\eqref{eq:WardJc}, every term cancels pairwise, resulting in
\begin{equation}
q_{0\mu}\cJ_{1\bar2|3}^\mu=0
\qquad\Longleftrightarrow\qquad
K_\mu\cJ_{1\bar2|3}^\mu=0.
\label{eq:rank3WardFinal}
\end{equation}

\subsection{Three-particle photon current after the Ward identity}
The full 3-point BG current is
\begin{equation}
\Ahat_{1\bar2|3}^\mu
=
\left[
\frac{P_T^{\mu\nu}(K)}{\DA(K^2)}
+\frac{\xi P_L^{\mu\nu}(K)}{K^2}
\right]
\cJ_{1\bar2|3,\nu}.
\label{eq:A123Projector}
\end{equation}
However, \eqref{eq:rank3WardFinal} implies
\begin{equation}
P_L^{\mu\nu}(K)\cJ_{1\bar2|3,\nu}=0,
\qquad
P_T^{\mu\nu}(K)\cJ_{1\bar2|3,\nu}
=\cJ_{1\bar2|3}^\mu.
\end{equation}
Therefore
\begin{equation}
\Ahat_{1\bar2|3}^\mu
=
\frac{\cJ_{1\bar2|3}^\mu}
{K^2(1-K^2/M_A^2)}.
\label{eq:A123HD}
\end{equation}
This is the same compact expression obtained in the earlier calculation, but
we have now shown precisely why the longitudinal part of the exact propagator
drops out.

\subsection{Perturbiner expansion in the higher derivative theory}
\label{sec:perturbiner}
Let us now derive all-rank recursions directly from the classical field equations i.e, the perturbiner. We then show that the above lower ranked currents are correctly produced in this prescription.

\subsubsection{Perturbiner ansatz}

We introduce some commuting bookkeeping variables $t_i$ for the elementary external
seeds.  For a sector with one scalar seed $1$ and photon labels $P$, we write
\begin{equation}
\phihat(x)
=
\sum_{P}(-1)^{|P|}\,
\Phi_{1|P}\,t_1t_P\,
 e^{-\ii k_{1P}\cdot x}
+\cdots .
\label{eq:phiPert}
\end{equation}
The factor $(-1)^{|P|}$ is a convenient BG convention. With it, the
perturbiner coefficients reproduce directly the diagrammatic currents of
sec.~\ref{sec:scalarDiagrams}, for which sequential one-photon insertions
come with positive sign and an $r$-photon local insertion carries
$(-1)^{r+1}$ in the recursion.
Similarly, in the sector with one scalar-antiscalar pair and additional
photon labels $P$, write
\begin{equation}
\Ahat^\mu(x)
=
\sum_i\eps_i^\mu t_i e^{-\ii k_i\cdot x}
+
\sum_P(-1)^{|P|}\,
\Ahat_{1\bar2|P}^\mu
t_1t_{\bar2}t_P
e^{-\ii k_{1\bar2P}\cdot x}
+\cdots .
\label{eq:APert}
\end{equation}
The elementary on-shell gauge legs satisfy
\begin{equation}
k_i\cdot\eps_i=0,
\qquad
\DA(k_i^2)=0,
\end{equation}
which implies either
\begin{equation}
k_i^2=0
\qquad\text{or}\qquad
k_i^2=M_A^2.
\end{equation}

We use
\begin{equation}
\Kphi(z)=z-\frac{z^2}{M_\phi^2}-m^2,
\qquad
\Dhat_\mu=\partial_\mu+ig\Ahat_\mu .
\end{equation}
The full nonlinear scalar equation of motion is
\begin{equation}
\Kphi(-\Dhat^2)\phihat=0,
\qquad
-\Dhat^2=-\Box-ig\Bigl(2\Ahat\!\cdot\!\partial+\partial\!\cdot\!\Ahat\Bigr)
+g^2\Ahat^2 .
\label{eq:scalarEOM}
\end{equation}
The full photon equation of motion is
\begin{equation}
\left(1+\frac{\Box}{M_A^2}\right)\partial_\nu \widehat F^{\nu\mu}
+\frac{1}{\xi}\,\partial^\mu(\partial\!\cdot\!\Ahat)
=J_{\rm mat}^{\mu}[\phihat,\phibarhat,\Ahat].
\label{eq:photonEOM}
\end{equation}

We now insert the three perturbiner expansions
\begin{equation}
\phihat(x)=\sum_P\Phi_P\,t_Pe^{ik_P\cdot x},\qquad
\phibarhat(x)=\sum_P\bar\Phi_P\,t_Pe^{ik_P\cdot x},\qquad
\Ahat^\mu(x)=\sum_P A_P^\mu\,t_Pe^{ik_P\cdot x}.
\label{eq:pertAnsatz}
\end{equation}
Here $t_P\equiv\prod_{i\in P}t_i$ with commuting bookkeeping variables $t_i$, and $k_P\equiv\sum_{i\in P}k_i$.  The coefficient of every independent monomial $t_Pe^{ik_P\cdot x}$ is then set to zero.

\subsection*{Scalar-current recursion}

Let $\Gamma_r$ denote the fully Bose-symmetrized scalar vertex with one incoming scalar current, one outgoing scalar root, and $r$ photon currents, as generated directly by \eqref{eq:scalarEOM}.  Because $X$ contains at most two powers of $\Ahat$ and the equation contains $X^2$, one has $r\leq4$.  For a scalar current with one distinguished on-shell scalar $1$ and a photon word $P$, extraction of the coefficient of $t_1t_Pe^{ik_{1P}\cdot x}$ gives
\begin{equation}
\Kphi(k_{1P}^2)\,\Phi_{1|P}
=
\sum_{r=1}^{\min(4,|P|)}
\sum_{\substack{P=Q\sqcup R_1\sqcup\cdots\sqcup R_r\\ R_j\neq\varnothing}}^{\rm unord.}
\Gamma_r\!\left(k_{1P};k_{1Q};R_1,\ldots,R_r\right)
\Phi_{1|Q}\prod_{j=1}^{r}A_{R_j}.
\label{eq:scalarRecursionImplicit}
\end{equation}
Hence
\begin{equation}
\Phi_{1|P}
=
\frac{1}{\Kphi(k_{1P}^2)}
\sum_{r=1}^{\min(4,|P|)}
\sum_{\substack{P=Q\sqcup R_1\sqcup\cdots\sqcup R_r\\ R_j\neq\varnothing}}^{\rm unord.}
\Gamma_r\,\Phi_{1|Q}\prod_{j=1}^{r}A_{R_j}.
\label{eq:scalarRecursion}
\end{equation}
The case $Q=\varnothing$ means $\Phi_{1|\varnothing}\equiv\Phi_1$.

\subsection*{Photon-current recursion}

The coefficient of a word with on-shell scalar leg $1$, on-shell antiscalar leg $\bar2$, and photon word $P$ gives
\begin{equation}
\left[
\DA(K^2)\PT^{\mu\nu}(K)
+\frac{K^2}{\xi}\PL^{\mu\nu}(K)
\right]A_{1\bar2|P,\nu}
=\cJ_{1\bar2|P}^{\mu},
\qquad K=k_1+k_2+k_P .
\label{eq:photonRecursionProjector}
\end{equation}
The matter source is obtained from the same scalar vertices.  Since one photon leg is the off-shell root, at most three additional photon currents can appear,
\begin{equation}
\cJ_{1\bar2|P}^{\mu}
=
\sum_{r=0}^{\min(3,|P|)}
\sum_{\substack{P=Q\sqcup R\sqcup S_1\sqcup\cdots\sqcup S_r\\ S_j\neq\varnothing}}^{\rm unord.}
\bar\Phi_{\bar2|R}\,
\Gamma_{r+1}^{\mu\nu_1\cdots\nu_r}
\Phi_{1|Q}
\prod_{j=1}^{r}A_{S_j,\nu_j}.
\label{eq:photonSourceRecursion}
\end{equation}
Therefore the photon BG current is
\begin{equation}
A_{1\bar2|P}^{\mu}
=
\left[
\frac{\PT^{\mu\nu}(K)}{\DA(K^2)}
+\frac{\xi\,\PL^{\mu\nu}(K)}{K^2}
\right]\cJ_{1\bar2|P,\nu}.
\label{eq:photonRecursion}
\end{equation}
Equations~\eqref{eq:scalarRecursion} and \eqref{eq:photonRecursion} are the coupled Berends--Giele recursions. The scalar current depends on lower-rank photon currents, while the photon current depends on lower-rank scalar, antiscalar, and photon currents.

\subsection{Explicit perturbiner check: scalar currents}
\label{sec:scalarPertCheck}

\subsubsection{Rank two}

In this case, we retain only
\begin{equation}
\phihat
=t_1e^{-\ii k_1x}
-t_1t_2\Phi_{1|2}e^{-\ii k_{12}x}+\cdots,
\end{equation}
\begin{equation}
\Ahat_\mu
=t_2\eps_{2\mu}e^{-\ii k_2x}+\cdots.
\end{equation}
At order $t_1t_2$, the scalar equation gives
\begin{equation}
-D_{12}\Phi_{1|2}
+g\cV_1^\mu(k_{12},k_1)\eps_{2\mu}=0.
\end{equation}
Hence
\begin{equation}
\Phi_{1|2}
=\frac{g\cV_1^\mu(k_{12},k_1)\eps_{2\mu}}{D_{12}},
\end{equation}
which is exactly \eqref{eq:Phi12Diagram}.

\subsubsection{Rank three}

For rank three, we keep
\begin{align}
\phihat
={}&t_1e^{-\ii k_1x}
-t_1t_2\Phi_{1|2}e^{-\ii k_{12}x}
-t_1t_3\Phi_{1|3}e^{-\ii k_{13}x}
\nonumber\\
&+t_1t_2t_3\Phi_{1|23}e^{-\ii k_{123}x}+\cdots,
\end{align}
and
\begin{equation}
\Ahat_\mu
=t_2\eps_{2\mu}e^{-\ii k_2x}
+t_3\eps_{3\mu}e^{-\ii k_3x}+\cdots.
\end{equation}
The coefficient of $t_1t_2t_3e^{-\ii k_{123}x}$ in the scalar equation is
\begin{align}
0={}&
D_{123}\Phi_{1|23}
-g\cV_1(k_{123},k_{13};2)\Phi_{1|3}
-g\cV_1(k_{123},k_{12};3)\Phi_{1|2}
\nonumber\\
&+g^2\cV_2(k_{123},k_1;2,3).
\end{align}
Therefore
\begin{equation}
\begin{aligned}
\Phi_{1|23}
=\frac{1}{D_{123}}\Big[
&g\cV_1(k_{123},k_{13};2)\Phi_{1|3}
+g\cV_1(k_{123},k_{12};3)\Phi_{1|2}
\\
&-g^2\cV_2(k_{123},k_1;2,3)
\Big].
\end{aligned}
\label{eq:Phi123Pert}
\end{equation}
Substituting the already-derived rank-two currents gives precisely the two
exchange graphs \eqref{eq:Phi123a}, \eqref{eq:Phi123b} and the contact graph
\eqref{eq:Phi123c}.  Hence
\begin{equation}
\Phi_{1|23}^{\rm perturbiner}
=
\Phi_{1|23}^{\rm diagrams}.
\end{equation}

\subsection{Explicit perturbiner check: photon currents}
\label{sec:photonPertCheck}

\subsubsection{Rank two}

Here we keep the scalar and antiscalar perturbiner expansions
\begin{equation}
\phihat=t_1e^{-\ii k_1x}+\cdots,
\qquad
\phihat^*=t_{\bar2}e^{-\ii k_2x}+\cdots,
\end{equation}
and the composite photon term
\begin{equation}
\Ahat^\mu
=t_1t_{\bar2}\Ahat_{1\bar2}^\mu e^{-\ii Kx}+\cdots,
\qquad K=k_1+k_2.
\end{equation}
At order $t_1t_{\bar2}$, the equation of motion gives
\begin{equation}
\left[
\DA(K^2)P_T^{\mu\nu}(K)
+\frac{K^2}{\xi}P_L^{\mu\nu}(K)
\right]
\Ahat_{1\bar2,\nu}
=\cJ_{1\bar2}^\mu,
\end{equation}
where
\begin{equation}
\cJ_{1\bar2}^\mu
=g(k_1-k_2)^\mu C(k_1^2,k_2^2).
\end{equation}
Inverting the quadratic operator gives
\begin{equation}
\Ahat_{1\bar2}^\mu
=
\left[
\frac{P_T^{\mu\nu}}{\DA(K^2)}
+\frac{\xi P_L^{\mu\nu}}{K^2}
\right]
\cJ_{1\bar2,\nu}.
\end{equation}
But \eqref{eq:rank2Ward} shows that the source is transverse for
on-shell scalar seeds.  Hence
\begin{equation}
\Ahat_{1\bar2}^\mu
=
\frac{g(k_1-k_2)^\mu C(k_1^2,k_2^2)}
{K^2(1-K^2/M_A^2)},
\end{equation}
which is exactly the diagrammatic result \eqref{eq:A12HD}.

\subsubsection{Rank three}

We now include one elementary photon seed $3$ and the relevant scalar sub-currents,
\begin{align}
\phihat
={}&t_1e^{-\ii k_1x}
-t_1t_3\Phi_{1|3}e^{-\ii k_{13}x}+\cdots,
\\
\phihat^*
={}&t_{\bar2}e^{-\ii k_2x}
-t_{\bar2}t_3\bar\Phi_{2|3}e^{-\ii k_{23}x}+\cdots,
\\
\Ahat^\mu
={}&t_3\eps_3^\mu e^{-\ii k_3x}
-t_1t_{\bar2}t_3\Ahat_{1\bar2|3}^\mu e^{-\ii Kx}
+\cdots.
\end{align}
The coefficient of $t_1t_{\bar2}t_3e^{-\ii Kx}$ in the gauge equation is
\begin{equation}
\begin{aligned}
&\left[
\DA(K^2)P_T^{\mu\nu}(K)
+\frac{K^2}{\xi}P_L^{\mu\nu}(K)
\right]
\Ahat_{1\bar2|3,\nu}
\\
&\qquad=
g^2\eps_{3\rho}\Big[
\cV_1^\mu(p',r_3)\Delta_{13}\cV_1^\rho(r_3,p)
+\cV_1^\rho(p',r_0)\Delta_{23}\cV_1^\mu(r_0,p)
-\cV_2^{\mu\rho}(p',p;q_0,k_3)
\Big].
\end{aligned}
\label{eq:A123PertEquation}
\end{equation}
The right-hand side is precisely the diagrammatic source
$\cJ_{1\bar2|3}^\mu$. Thus, inversion gives
\begin{equation}
\Ahat_{1\bar2|3}^\mu
=
\left[
\frac{P_T^{\mu\nu}(K)}{\DA(K^2)}
+\frac{\xi P_L^{\mu\nu}(K)}{K^2}
\right]
\cJ_{1\bar2|3,\nu}.
\end{equation}
The explicit calculation in Sec.~\ref{sec:rank3Ward} proves
$K_\nu\cJ_{1\bar2|3}^\nu=0$, so
\begin{equation}
\Ahat_{1\bar2|3}^{\mu,\rm perturbiner}
=
\frac{\cJ_{1\bar2|3}^\mu}{\DA(K^2)}
=
\Ahat_{1\bar2|3}^{\mu,\rm diagrams}.
\end{equation}
The perturbiner therefore reproduces the full projector calculation and its
Ward-identity reduction exactly through rank three.
\section{Berends--Giele currents in higher-derivative Yang--Mills}
\label{sec:HDYM_BG}

We consider the higher-derivative Yang--Mills theory
\begin{equation}
\mathcal L_{\rm HDYM}
=
-\frac14 F^a_{\mu\nu}F^{a\mu\nu}
+\frac{1}{2M^2}
(D_\mu F^{a\mu\nu})(D^\lambda F^a{}_{\lambda\nu}),
\label{eq:HDYM_L}
\end{equation}
with
\begin{align}
F^a_{\mu\nu}
&=
\partial_\mu A^a_\nu-\partial_\nu A^a_\mu
+g f^{abc}A^b_\mu A^c_\nu,
\\
(D_\rho F_{\mu\nu})^a
&=
\partial_\rho F^a_{\mu\nu}
+g f^{abc}A^b_\rho F^c_{\mu\nu}.
\end{align}
Here, $M$ is the higher derivative scale of the gluon, interpreted as the mass of the Merlin mode. From the above, both the propagator and the interactions are modified. While the propagator develops a new pole, see below, the interactions extend to sextic ones instead of only cubic or quartic. However, the structures of the interactions are highly constrained, and this will show up in the proof of the all on-shell massless gluon BG current in this theory. The higher derivatives in the Lagrangian result in a propagator which is quartic in the momentum, and thus in loop calculations, the divergence is tamed and it results in a better UV behavior.
\subsection{Generalized Feynman gauge and propagator}
We choose the higher-derivative covariant gauge-fixing term
\begin{equation}
\mathcal L_{\rm gf}
=
-\frac{1}{2\xi}
(\partial_\mu A^{a\mu})
\left(1+\frac{\Box}{M^2}\right)
(\partial_\nu A^{a\nu}) .
\end{equation}
The momentum-space propagator in the gauge $\xi=1$ is then
\begin{equation}
D_{\mu\nu}^{ab}(p)
=
-\frac{i\delta^{ab}\eta_{\mu\nu}}
{p^2(1-p^2/M^2)}
=
-i\delta^{ab}\eta_{\mu\nu}
\left(
\frac{1}{p^2}
-\frac{1}{p^2-M^2}
\right).
\end{equation}
Thus the same higher-derivative gauge field contains a massless gluon pole at
$K^2=0$ and a massive pole at $K^2=M^2$.  Following the standard terminology,
we refer to the massive vector mode as a Merlin.  For an ordered word
$P=i_1i_2\cdots i_m$ we use
\begin{equation}
k_P\equiv\sum_{i\in P}k_i,
\qquad
s_P\equiv k_P^2,
\qquad
\Delta_P\equiv\frac{1}{s_P(1-s_P/M^2)}.
\label{eq:HDYM_word_notation}
\end{equation}
An overall factor $g^{|P|-1}$ is stripped from the color-ordered BG current.

\subsection{Cubic Feynman rule}
\label{subsec:HDYM_cubic_rule}

As to what follows, we will derive the BG current explicitly upto 3 points and thus we just need the cubic Feynman rule. The general recursion will of course need vertices up to six gauge fields but it will not be explicit in our construction. So let us write the exact cubic vertex. With all momenta incoming,
$k+p+q=0$, the cubic vertex is written as
\begin{equation}
V^{abc}_{\mu\nu\rho}(k,p,q)
=
g f^{abc}\,\widehat V_{\mu\nu\rho}(k,p,q),
\end{equation}
where
\begin{align}
\widehat V_{\mu\nu\rho}(k,p,q)
={}&
\left(1-\frac{k^2+p^2+q^2}{M^2}\right)
V^{\rm YM}_{\mu\nu\rho}(k,p,q)
\nonumber\\
&+\frac1{M^2}\Big\{
 k_\mu\big[(q_\nu-k_\nu)k_\rho
 +k\cdot(p-q)\eta_{\nu\rho}
 +(k_\rho-p_\rho)k_\nu\big]
\nonumber\\
&\hspace{17mm}
+p_\nu\big[(p_\mu-q_\mu)p_\rho
+p\cdot(q-k)\eta_{\rho\mu}
+(k_\rho-p_\rho)p_\mu\big]
\nonumber\\
&\hspace{17mm}
+q_\rho\big[(q_\nu-k_\nu)q_\mu
+q\cdot(k-p)\eta_{\mu\nu}
+(p_\mu-q_\mu)q_\nu\big]
\Big\},
\label{eq:HDYM_full_cubic_vertex}
\end{align}
with the ordinary Yang--Mills tensor
\begin{equation}
V^{\rm YM}_{\mu\nu\rho}(k,p,q)
=
\eta_{\mu\nu}(k-p)_\rho
+\eta_{\nu\rho}(p-q)_\mu
+\eta_{\rho\mu}(q-k)_\nu.
\label{eq:HDYM_YM_cubic_tensor}
\end{equation} 
The reduced formula used below is obtained only after contracting two legs with transverse
currents.
\subsubsection{Quartic contact vertex}

The three point BG current also receives a contribution from the quartic
higher-derivative Yang-Mills vertex.  For the color ordering $(1,2,3)$,
with off-shell momentum
\begin{equation}
K=k_1+k_2+k_3,
\qquad
s=K^2,
\end{equation}
we write the contracted quartic vertex as
\begin{equation}
\mathbb V_4^\mu(1,2,3)
=
\mathbb V_{4,\rm YM}^\mu(1,2,3)
+\frac{1}{M^2}\,
\mathbb V_{4,\rm HD}^\mu(1,2,3).
\end{equation}
The ordinary Yang-Mills contribution is
\begin{equation}
\mathbb V_{4,\rm YM}^\mu(1,2,3)
=
2(\varepsilon_1\!\cdot\!\varepsilon_3)\varepsilon_2^\mu
-(\varepsilon_1\!\cdot\!\varepsilon_2)\varepsilon_3^\mu
-(\varepsilon_2\!\cdot\!\varepsilon_3)\varepsilon_1^\mu .
\end{equation}

For three on-shell massless external gluons,
$k_i^2=0$ and $k_i\!\cdot\!\varepsilon_i=0$, the higher-derivative
contact term can be written as
\begin{align}
\mathbb V_{4,\rm HD}^\mu(1,2,3)
={}&
\mathbb V_{\rm YM}^\mu(K_{12},k_3;n_{12},\varepsilon_3)
+\mathbb V_{\rm YM}^\mu(k_1,K_{23};\varepsilon_1,n_{23})
\nonumber\\
&-K^\mu(n_{12}\!\cdot\!\varepsilon_3)
+K^\mu(\varepsilon_1\!\cdot n_{23})
-s\,\mathbb V_{4,\rm YM}^\mu(1,2,3),
\label{eq:HDYM_quartic_contact}
\end{align}
where
\begin{align}
n_{12}^\mu
&=
(\varepsilon_1\!\cdot\!\varepsilon_2)(k_1-k_2)^\mu
+2(k_2\!\cdot\!\varepsilon_1)\varepsilon_2^\mu
-2(k_1\!\cdot\!\varepsilon_2)\varepsilon_1^\mu,
\\
n_{23}^\mu
&=
(\varepsilon_2\!\cdot\!\varepsilon_3)(k_2-k_3)^\mu
+2(k_3\!\cdot\!\varepsilon_2)\varepsilon_3^\mu
-2(k_2\!\cdot\!\varepsilon_3)\varepsilon_2^\mu ,
\end{align}
and
\begin{equation}
\mathbb V_{\rm YM}^\mu(P,Q;X,Y)
=
(X\!\cdot\!Y)(P-Q)^\mu
+\big[(P+2Q)\!\cdot X\big]Y^\mu
-\big[(2P+Q)\!\cdot Y\big]X^\mu .
\end{equation}

\subsection{Cubic current map and perturbiner recursion}
\label{subsec:HDYM_setup}

For later use it is convenient to contract the two daughter legs of the cubic
vertex from the beginning.  Let
\begin{equation}
K=P+Q,
\qquad
P\cdot X=0,
\qquad
Q\cdot Y=0,
\end{equation}
where $X^\mu(P)$ and $Y^\mu(Q)$ are transverse daughter currents.  The ordinary
Yang--Mills cubic current map is
\begin{align}
\mathbb V_{\rm YM}^\mu(P,Q;X,Y)
={}&
(X\cdot Y)(P-Q)^\mu
+\big[(P+2Q)\cdot X\big]Y^\mu
\nonumber\\
&-\big[(2P+Q)\cdot Y\big]X^\mu.
\label{eq:HDYM_YMmap}
\end{align}
The cubic vertex of the higher-derivative theory reduces, for transverse daughter
currents, to
\begin{align}
\mathbb V_{\rm HD}^\mu(P,Q;X,Y)
={}&
\left(1-\frac{K^2+P^2+Q^2}{M^2}\right)
\mathbb V_{\rm YM}^\mu(P,Q;X,Y)
\nonumber\\
&+
\frac{P^2-Q^2}{M^2}
K^\mu(X\cdot Y).
\label{eq:HDYM_cubic_map}
\end{align}
This compact formula is sufficient for all low-rank calculations below.  Notice
that the higher-derivative cubic vertex is genuinely different from the usual
Yang--Mills vertex. As we will use, the equality of the massless branch BG currents with their Yang-Mills
counterparts will therefore arise from a cancellation between the modified
vertex numerator and the modified propagator.
The matrix-valued perturbiner is
\begin{equation}
A^\mu(x)
=
\sum_{P\neq\emptyset}
J_P^\mu T^P e^{ik_P\cdot x},
\qquad
T^P=T^{a_{i_1}}\cdots T^{a_{i_m}},
\label{eq:HDYM_perturbiner}
\end{equation}
with one-particle seeds $J_i^\mu=\varepsilon_i^\mu$.  Products of fields induce
ordered deconcatenations of words.  Since the interaction
$(D_\mu F^{\mu\nu})^2$ contains vertices up to six gauge fields, the general
color-ordered recursion may be written as
\begin{align}
J_P^\mu
=
\Delta_P\Bigg[{}
&\sum_{P=QR}
\mathbb V_3^\mu(Q,R;J_Q,J_R)
+\sum_{P=QRS}
\mathbb V_4^\mu(Q,R,S;J_Q,J_R,J_S)
\nonumber\\
&+\sum_{P=QRST}
\mathbb V_5^\mu(Q,R,S,T;J_Q,J_R,J_S,J_T)
\nonumber\\
&+\sum_{P=QRSTU}
\mathbb V_6^\mu(Q,R,S,T,U;J_Q,J_R,J_S,J_T,J_U)
\Bigg],
\label{eq:HDYM_allrank_recursion}
\end{align}
where every sum is over nonempty ordered deconcatenations.  The kernels
$\mathbb V_r$ are obtained by differentiating the corresponding interaction
terms of \eqref{eq:HDYM_L}.  At rank two only $\mathbb V_3$ contributes while
at rank three only $\mathbb V_3$ and $\mathbb V_4$ contribute.

\subsection{Massless branch BG currents from Feynman diagrams}
\label{subsec:HDYM_massless_feynman}
We first consider all external on-shell legs to be ordinary massless gluons,
\begin{equation}
k_i^2=0,
\qquad
k_i\cdot\varepsilon_i=0.
\label{eq:HDYM_massless_seed}
\end{equation}
The first two nontrivial currents already display the mechanism that will later
become an all-rank statement.

\subsubsection{Two-particle current}

For the ordered word $12$ there is only one cubic diagram.  With
$K_{12}=k_1+k_2$ and $s_{12}=K_{12}^2$, \eqref{eq:HDYM_cubic_map} gives
\begin{equation}
\mathbb V_{\rm HD}^\mu(k_1,k_2;\varepsilon_1,\varepsilon_2)
=
\left(1-\frac{s_{12}}{M^2}\right)
\mathbb V_{\rm YM}^\mu(k_1,k_2;\varepsilon_1,\varepsilon_2),
\label{eq:HDYM_rank2_vertex_reduction}
\end{equation}
because $k_1^2=k_2^2=0$. The current is therefore given by
\begin{align}
J_{12}^\mu
&=
\frac{1}{s_{12}(1-s_{12}/M^2)}
\left(1-\frac{s_{12}}{M^2}\right)
\mathbb V_{\rm YM}^\mu(1,2)
\nonumber\\
&=
\frac{1}{s_{12}}\mathbb V_{\rm YM}^\mu(1,2).
\label{eq:HDYM_J12_equals_YM}
\end{align}
Using \eqref{eq:HDYM_YMmap}, this is
\begin{equation}
J_{12}^\mu
=
\frac1{s_{12}}
\left[
(\varepsilon_1\cdot\varepsilon_2)(k_1-k_2)^\mu
+2(k_2\cdot\varepsilon_1)\varepsilon_2^\mu
-2(k_1\cdot\varepsilon_2)\varepsilon_1^\mu
\right].
\label{eq:HDYM_J12_explicit}
\end{equation}
Thus the two-particle higher-derivative current is exactly the usual
Yang-Mills current. To arrive at this equality, the factor
$1-s_{12}/M^2$ in the cubic numerator has cancelled the massive factor in the higher derivative off-shell propagator. Consequently, the current has only the massless pole $1/s_{12}$ and no pole at $s_{12}=M^2$.
The current satisfies
\begin{equation}
K_{12}\cdot J_{12}=0,
\label{eq:HDYM_J12_transverse}
\end{equation}

\subsection{Three-particle BG current and its massless-branch reduction}
\label{subsec:HDYM_J123}

We now construct the three-particle Berends--Giele current in
higher-derivative Yang-Mills.  The external on-shell legs are ordinary
massless gluons,
\begin{equation}
k_i^2=0,
\qquad
k_i\cdot\varepsilon_i=0,
\qquad i=1,2,3.
\end{equation}
For an ordered word $P$ we use
\begin{equation}
K_P=\sum_{i\in P}k_i,
\qquad
s_P=K_P^2,
\qquad
\Delta_P^{\rm HD}
=
\frac{1}{s_P(1-s_P/M^2)}.
\end{equation}

For the ordered word $(123)$ there are two cubic exchange diagrams and one
quartic contact diagram.  Defining
\begin{equation}
K\equiv K_{123}=k_1+k_2+k_3,
\qquad
s\equiv K^2,
\end{equation}
the complete Feynman-diagram expression is
\begin{align}
J_{123}^{\mu,{\rm HD}}
=
\Delta_{123}^{\rm HD}
\Big[
&
\mathbb V_{\rm HD}^{\mu}
\Big(
K_{12},k_3;
\Delta_{12}^{\rm HD}\,
\mathbb V_{\rm HD}(k_1,k_2;
\varepsilon_1,\varepsilon_2),
\varepsilon_3
\Big)
\nonumber\\
&+
\mathbb V_{\rm HD}^{\mu}
\Big(
k_1,K_{23};
\varepsilon_1,
\Delta_{23}^{\rm HD}\,
\mathbb V_{\rm HD}(k_2,k_3;
\varepsilon_2,\varepsilon_3)
\Big)
\nonumber\\
&+
\mathbb V_{4,{\rm HDYM}}^\mu(1,2,3)
\Big].
\label{eq:HDYM_J123_full}
\end{align}
Thus $\Delta_{12}^{\rm HD}$ and $\Delta_{23}^{\rm HD}$ are the internal
propagators of the two exchange diagrams, while
$\Delta_{123}^{\rm HD}$ is the propagator of the off-shell leg.

It is convenient to introduce the rank-two HDYM currents
\begin{align}
J_{12}^{\mu,{\rm HD}}
&=
\Delta_{12}^{\rm HD}\,
\mathbb V_{\rm HD}^{\mu}
(k_1,k_2;\varepsilon_1,\varepsilon_2),
\\
J_{23}^{\mu,{\rm HD}}
&=
\Delta_{23}^{\rm HD}\,
\mathbb V_{\rm HD}^{\mu}
(k_2,k_3;\varepsilon_2,\varepsilon_3).
\end{align}
For two massless external gluons the cubic HD vertex satisfies
\begin{equation}
\mathbb V_{\rm HD}^{\mu}
(k_i,k_j;\varepsilon_i,\varepsilon_j)
=
\left(1-\frac{s_{ij}}{M^2}\right)
\mathbb V_{\rm YM}^{\mu}
(k_i,k_j;\varepsilon_i,\varepsilon_j),
\end{equation}
and therefore
\begin{equation}
J_{ij}^{\mu,{\rm HD}}
=
\frac{
\mathbb V_{\rm YM}^{\mu}
(k_i,k_j;\varepsilon_i,\varepsilon_j)
}{s_{ij}}.
\label{eq:HDYM_rank2_reduction}
\end{equation}
Hence the massive factor in the internal higher derivative propagator cancels against the
corresponding factor in the cubic HD vertex. 
Using the lower-rank currents, \eqref{eq:HDYM_J123_full} becomes
\begin{align}
J_{123}^{\mu,{\rm HD}}
=
\Delta_{123}^{\rm HD}
\Big[
&
\mathbb V_{\rm HD}^{\mu}
(K_{12},k_3;J_{12}^{\rm HD},\varepsilon_3)
+
\mathbb V_{\rm HD}^{\mu}
(k_1,K_{23};\varepsilon_1,J_{23}^{\rm HD})
\nonumber\\
&+
\mathbb V_{4,{\rm HDYM}}^\mu(1,2,3)
\Big].
\label{eq:HDYM_J123_compact}
\end{align}

For transverse daughter currents the cubic HDYM map can be written as
\begin{equation}
\mathbb V_{\rm HD}^{\mu}(P,Q;X,Y)
=
\left(
1-\frac{K^2+P^2+Q^2}{M^2}
\right)
\mathbb V_{\rm YM}^{\mu}(P,Q;X,Y)
+
\frac{P^2-Q^2}{M^2}
K^\mu(X\cdot Y),
\label{eq:HDYM_cubic_rank3}
\end{equation}
where $K=P+Q$.
Let us define
\begin{align}
\mathcal A^\mu
&\equiv
\mathbb V_{\rm YM}^{\mu}
(K_{12},k_3;J_{12}^{\rm HD},\varepsilon_3),
\\
\mathcal B^\mu
&\equiv
\mathbb V_{\rm YM}^{\mu}
(k_1,K_{23};\varepsilon_1,J_{23}^{\rm HD}),
\end{align}
where the internal propagators are already contained in
$J_{12}^{\rm HD}$ and $J_{23}^{\rm HD}$.  Applying \eqref{eq:HDYM_cubic_rank3} gives the two higher derivative exchange contributions
\begin{align}
\mathcal E_{12|3}^{\mu}
&\equiv
\mathbb V_{\rm HD}^{\mu}
(K_{12},k_3;J_{12}^{\rm HD},\varepsilon_3)
\nonumber\\
&=
\left(
1-\frac{s+s_{12}}{M^2}
\right)\mathcal A^\mu
+
\frac{s_{12}}{M^2}
K^\mu
(J_{12}^{\rm HD}\cdot\varepsilon_3),
\label{eq:HDYM_E12_3}
\\
\mathcal E_{1|23}^{\mu}
&\equiv
\mathbb V_{\rm HD}^{\mu}
(k_1,K_{23};\varepsilon_1,J_{23}^{\rm HD})
\nonumber\\
&=
\left(
1-\frac{s+s_{23}}{M^2}
\right)\mathcal B^\mu
-
\frac{s_{23}}{M^2}
K^\mu
(\varepsilon_1\cdot J_{23}^{\rm HD}).
\label{eq:HDYM_E1_23}
\end{align}

The quartic contact term is separated into its ordinary Yang--Mills and
higher-derivative pieces,
\begin{equation}
\mathbb V_{4,{\rm HDYM}}^\mu(1,2,3)
=
\mathcal C^\mu
+
\frac{1}{M^2}
\mathbb V_{4,{\rm HD}}^\mu(1,2,3),
\end{equation}
where
\begin{equation}
\mathcal C^\mu
\equiv
\mathbb V_{4,{\rm YM}}^\mu(1,2,3)
=
2(\varepsilon_1\cdot\varepsilon_3)\varepsilon_2^\mu
-(\varepsilon_1\cdot\varepsilon_2)\varepsilon_3^\mu
-(\varepsilon_2\cdot\varepsilon_3)\varepsilon_1^\mu.
\label{eq:HDYM_C_contact}
\end{equation}

For three massless external gluons, direct evaluation of the
higher-derivative quartic interaction gives
\begin{align}
\mathbb V_{4,{\rm HD}}^\mu(1,2,3)
={}&
s_{12}\mathcal A^\mu
+s_{23}\mathcal B^\mu
-s_{12}K^\mu
(J_{12}^{\rm HD}\cdot\varepsilon_3)
\nonumber\\
&+
s_{23}K^\mu
(\varepsilon_1\cdot J_{23}^{\rm HD})
-s\,\mathcal C^\mu.
\label{eq:HDYM_V4HD_rank3}
\end{align}

The complete HDYM three-particle current is therefore
\begin{equation}
J_{123}^{\mu,{\rm HD}}
=
\frac{
\mathcal E_{12|3}^{\mu}
+\mathcal E_{1|23}^{\mu}
+\mathcal C^\mu
+M^{-2}\mathbb V_{4,{\rm HD}}^\mu(1,2,3)
}
{s(1-s/M^2)}.
\label{eq:HDYM_J123_before_cancel}
\end{equation}
Substituting \eqref{eq:HDYM_E12_3},
\eqref{eq:HDYM_E1_23}, and
\eqref{eq:HDYM_V4HD_rank3} into the numerator gives
\begin{align}
\mathcal N_{123,{\rm HD}}^\mu
={}&
\mathcal A^\mu+\mathcal B^\mu+\mathcal C^\mu
\nonumber\\
&+
\frac{1}{M^2}
\Big[
-(s+s_{12})\mathcal A^\mu
-(s+s_{23})\mathcal B^\mu
\nonumber\\
&\hspace{13mm}
+s_{12}K^\mu(J_{12}^{\rm HD}\cdot\varepsilon_3)
-s_{23}K^\mu(\varepsilon_1\cdot J_{23}^{\rm HD})
\nonumber\\
&\hspace{13mm}
+s_{12}\mathcal A^\mu
+s_{23}\mathcal B^\mu
-s_{12}K^\mu(J_{12}^{\rm HD}\cdot\varepsilon_3)
\nonumber\\
&\hspace{13mm}
+s_{23}K^\mu(\varepsilon_1\cdot J_{23}^{\rm HD})
-s\,\mathcal C^\mu
\Big].
\end{align}
All terms proportional to $s_{12}$ and $s_{23}$ cancel pairwise, leaving
\begin{equation}
\mathcal N_{123,{\rm HD}}^\mu
=
\left(1-\frac{s}{M^2}\right)
\left(
\mathcal A^\mu+\mathcal B^\mu+\mathcal C^\mu
\right).
\label{eq:HDYM_rank3_factorization}
\end{equation}
Hence
\begin{align}
J_{123}^{\mu,{\rm HD}}
&=
\frac{
(1-s/M^2)
(\mathcal A^\mu+\mathcal B^\mu+\mathcal C^\mu)
}
{s(1-s/M^2)}
\nonumber\\
&=
\frac{
\mathcal A^\mu+\mathcal B^\mu+\mathcal C^\mu
}{s}.
\label{eq:HDYM_J123_after_cancel}
\end{align}
However, the above is actually the usual YM 3 point BG current
\begin{equation}
J_{123}^{\mu,{\rm YM}}
=
\frac{
\mathcal A^\mu+\mathcal B^\mu+\mathcal C^\mu
}{s}.
\end{equation}
Therefore, for three massless on-shell external gluons,
\begin{equation}
J_{123}^{\mu,{\rm HD}}
=
J_{123}^{\mu,{\rm YM}}.
\label{eq:HDYM_J123_equals_YM}
\end{equation}

The equality is nontrivial.  The two exchange diagrams separately contain
explicit higher-derivative corrections proportional to $s_{12}/M^2$ and
$s_{23}/M^2$, and the off-shell leg initially carries the full HD
propagator $1/[s(1-s/M^2)]$.  Only after including the higher-derivative
quartic contact interaction do all internal higher-derivative corrections combine into the
overall factor $1-s/M^2$, which cancels the massive pole of the off-shell final
propagator. 
\subsection{Perturbiner reproduction of the low-rank currents}
\label{subsec:HDYM_perturbiner_lowrank}

The same results follow directly from the perturbiner expansion.  At word $12$
we retain
\begin{equation}
A^\mu(x)
=
\varepsilon_1^\mu T^{a_1}e^{ik_1x}
+\varepsilon_2^\mu T^{a_2}e^{ik_2x}
+J_{12}^\mu T^{a_1}T^{a_2}e^{ik_{12}x}
+\cdots.
\end{equation}
The coefficient of $T^{a_1}T^{a_2}e^{ik_{12}x}$ in the classical equation is
\begin{equation}
s_{12}\left(1-\frac{s_{12}}{M^2}\right)J_{12}^\mu
=
\mathbb V_{\rm HD}^\mu(k_1,k_2;\varepsilon_1,\varepsilon_2).
\end{equation}
Using \eqref{eq:HDYM_rank2_vertex_reduction} immediately reproduces \eqref{eq:HDYM_J12_equals_YM}.
For the ordered word ($123$) the perturbiner contains
\begin{align}
A^\mu(x)
={}&
\sum_{i=1}^{3}\varepsilon_i^\mu T^{a_i}e^{ik_i x}
+J_{12}^\mu T^{a_1}T^{a_2}e^{ik_{12}x}
+J_{23}^\mu T^{a_2}T^{a_3}e^{ik_{23}x}
\nonumber\\
&+J_{123}^\mu T^{a_1}T^{a_2}T^{a_3}e^{ik_{123}x}
+\cdots.
\end{align}
There are exactly three source terms: the deconcatenations $12|3$ and $1|23$
from the cubic interaction, and $1|2|3$ from the quartic interaction.  Thus
\begin{align}
s_{123}\left(1-\frac{s_{123}}{M^2}\right)J_{123}^\mu
={}&
\mathbb V_{\rm HD}^\mu(K_{12},k_3;J_{12},\varepsilon_3)
\nonumber\\
&+
\mathbb V_{\rm HD}^\mu(k_1,K_{23};\varepsilon_1,J_{23})
+\mathbb V_4^\mu(1,2,3),
\label{eq:HDYM_perturbiner_rank3}
\end{align}
which is precisely the Feynman-diagram expression.  Hence the Feynman construction and the
perturbiner coefficient equations agree term by term at the first two
nontrivial ranks.  In a compact notation the correspondence is
\begin{align}
\text{Feynman diagrams:}\qquad
J_{12}
&=\Delta_{12}\,V_3(J_1,J_2),
\\
J_{123}
&=\Delta_{123}\big[
V_3(J_{12},J_3)+V_3(J_1,J_{23})+V_4(J_1,J_2,J_3)
\big],
\end{align}
while the perturbiner coefficients obey
\begin{align}
[12]:\qquad
s_{12}(1-s_{12}/M^2)J_{12}
&=V_3(J_1,J_2),
\\
[123]:\qquad
s_{123}(1-s_{123}/M^2)J_{123}
&=V_3(J_{12},J_3)+V_3(J_1,J_{23})
\nonumber\\
&\quad+V_4(J_1,J_2,J_3).
\end{align}
Thus the perturbiner
reconstructs the individual exchange and contact topologies generated by the
Feynman rules.

\subsection{All-rank equality of the massless HDYM and YM Berends--Giele currents}
\label{subsec:all-rank-hdym-bg}

We now prove by induction that, for purely massless external gluons, the
higher-derivative Berends-Giele current coincides with the
ordinary Yang-Mills current at arbitrary rank.

Let us decompose the Yang-Mills equation of motion as
\begin{equation}
E^\mu_{\rm YM}[A]
=
E^{(1)\mu}[A]
+
E^\mu_{\rm int}[A],
\end{equation}
where $E^{(1)}$ is the part linear in the gauge field and
$E_{\rm int}$ contains the nonlinear terms,
\begin{equation}
E_{\rm int}
=
gE^{(2)}+g^2E^{(3)}.
\end{equation}
The perturbiner expansion of the nonlinear part is written as
\begin{equation}
E^\mu_{\rm int}(x)
=
\sum_{P\neq\emptyset}
\mathcal S^\mu_{P,{\rm YM}}\,
T^P e^{iK_P\cdot x},
\end{equation}
so that, in the transverse sector,
\begin{equation}
E^{(1)\mu}_P
=
-s_PJ_P^\mu,
\qquad
s_P\equiv K_P^2,
\end{equation}
and the ordinary YM BG recursion is
\begin{equation}
s_PJ^\mu_{P,{\rm YM}}
=
\mathcal S^\mu_{P,{\rm YM}}.
\label{eq:ym-bg-source}
\end{equation}

The higher-derivative term is
\begin{equation}
\frac{1}{2M^2}
\left(E^{(1)}+E_{\rm int}\right)^2.
\end{equation}
The purely quadratic contribution
$\frac{1}{2M^2}(E^{(1)})^2$ is included in the higher derivative kinetic operator and
produces the propagator denominator
$s_P(1-s_P/M^2)$.  The remaining higher derivative interaction is therefore
\begin{equation}
\mathcal L_{{\rm HD,int}}
=
\frac{1}{M^2}
E_{\rm int}\cdot
\left(
E^{(1)}+\frac12E_{\rm int}
\right).
\label{eq:hd-int-factorized}
\end{equation}
Varying this term and integrating derivatives by parts gives
\begin{equation}
\delta S_{{\rm HD,int}}
=
\frac{1}{M^2}
\int d^Dx\,
\delta A\cdot
\left[
(E'_{\rm int})^\dagger
\left(E^{(1)}+E_{\rm int}\right)
+
(E^{(1)\prime})^\dagger E_{\rm int}
\right],
\label{eq:hd-int-var}
\end{equation}
where the prime denotes functional differentiation with respect to the gauge field.

We now assume that for every proper ordered subword $Q\subsetneq P$,
\begin{equation}
J^\mu_{Q,{\rm HD}}
=
J^\mu_{Q,{\rm YM}}.
\label{eq:induction-assumption}
\end{equation}
Then every lower-rank current obeys the YM BG equation,
\begin{equation}
\left(E^{(1)}+E_{\rm int}\right)_Q^\mu
=
-s_QJ_Q^\mu+\mathcal S^\mu_{Q,{\rm YM}}
=
0.
\label{eq:lower-ym-eom}
\end{equation}

Consider the first term in \eqref{eq:hd-int-var}.  Since
$E_{\rm int}$ starts at quadratic order in $A$, the operator
$E'_{\rm int}$ contains at least one field.  Hence, at total word $P$,
it can only multiply a coefficient of
$E^{(1)}+E_{\rm int}$ associated with a proper subword.  Therefore,
schematically,
\begin{equation}
\left[
(E'_{\rm int})^\dagger
\left(E^{(1)}+E_{\rm int}\right)
\right]_P
=
\sum_{Q\subsetneq P}
\mathcal K_{P,Q}
\left(
-s_QJ_Q+\mathcal S_{Q,{\rm YM}}
\right),
\end{equation}
where $\mathcal K_{P,Q}$ denotes the corresponding local YM kernel.
Using \eqref{eq:lower-ym-eom}, this contribution vanishes:
\begin{equation}
\left[
(E'_{\rm int})^\dagger
\left(E^{(1)}+E_{\rm int}\right)
\right]_P
=
0.
\label{eq:first-term-zero}
\end{equation}

The second term in \eqref{eq:hd-int-var} is simpler.  Since
$E^{(1)}$ is linear, $E^{(1)\prime}$ is just the free inverse-propagator
operator.  Acting on the word-$P$ coefficient of $E_{\rm int}$, which
carries total momentum $K_P$, it gives
\begin{equation}
\left[
(E^{(1)\prime})^\dagger E_{\rm int}
\right]_P^\mu
=
-s_P\,\mathcal S^\mu_{P,{\rm YM}}.
\label{eq:root-factor}
\end{equation}
Thus the full higher derivative interaction contributes
\begin{equation}
-\frac{s_P}{M^2}\,
\mathcal S^\mu_{P,{\rm YM}}
\end{equation}
to the BG source.  Adding the ordinary YM source therefore yields
\begin{equation}
\mathcal S^\mu_{P,{\rm HD}}
=
\left(
1-\frac{s_P}{M^2}
\right)
\mathcal S^\mu_{P,{\rm YM}}.
\label{eq:source-factorization}
\end{equation}

Finally,
\begin{align}
J^\mu_{P,{\rm HD}}
&=
\frac{
\mathcal S^\mu_{P,{\rm HD}}
}{
s_P(1-s_P/M^2)
}
\nonumber\\
&=
\frac{
(1-s_P/M^2)\mathcal S^\mu_{P,{\rm YM}}
}{
s_P(1-s_P/M^2)
}
=
\frac{
\mathcal S^\mu_{P,{\rm YM}}
}{s_P}
=
J^\mu_{P,{\rm YM}}.
\label{pole}
\end{align}
Hence the purely massless higher derivative YM Berends-Giele currents coincide with the
ordinary Yang-Mills currents at arbitrary rank.

\subsection{BG proof of the one-Merlin selection rule}
\label{subsec:HDYM_oneMerlin_selection}

The absence of the massive pole in \eqref{pole}
has an immediate amplitude consequence. To get a tree level amplitude, we amputate the off-shell leg of the BG current and project it to polarization.  For an ordinary gluon off-shell leg one takes
$s_P\to0$,
\begin{equation}
\mathcal A(g_1,\ldots,g_n,g)
\propto
\lim_{s_P\to0}s_P\,
\varepsilon_{g\mu}J_P^\mu,
\end{equation}
which reproduces the usual Yang-Mills amplitude.  If the same current is
instead projected onto the massive Merlin state, one must take the residue at
$s_P=M^2$,
\begin{equation}
\mathcal A(g_1,\ldots,g_n,B)
\propto
\lim_{s_P\to M^2}
(s_P-M^2)\,
\varepsilon_{B\mu}J_P^\mu.
\label{eq:HDYM_oneMerlin_residue}
\end{equation}
But \eqref{pole} is regular at $s_P=M^2$.
Therefore
\begin{equation}
\mathcal A(g_1,\ldots,g_n,B)=0.
\label{eq:HDYM_oneMerlin_zero}
\end{equation}
This gives a BG-current proof of the all-multiplicity one-Merlin selection rule.
The mechanism is especially transparent. Inspite of the massive pole being present in the final propagator of the BG current, it is cancelled by the numerator. Thus, when we amputate this off-shell leg, we multiply by the propagator factor and take the limit. Since there is no pole to cancel this factor, the limit results in the vanishing of the full amplitude.  Thus, the selection rule is encoded already at the level of the off-shell multiparticle current.

\subsection{BG currents with one on-shell Merlin}
\label{subsec:HDYM_oneMerlin_current}

A distinct multiparticle current is obtained when one of the on-shell leaves is
itself a Merlin particle. The simplest example has one ordinary massless gluon and one
Merlin,
\begin{equation}
k_1^2=0,
\qquad
k_2^2=M^2,
\qquad
k_i\cdot\varepsilon_i=0,
\end{equation}
with $K=k_1+k_2$.  Substitution into Eq.~\eqref{eq:HDYM_cubic_map} gives
\begin{align}
\mathbb V_{\rm HD}^\mu(A_1,B_2)
&=
-\frac{K^2}{M^2}
\mathbb V_{\rm YM}^\mu(A_1,B_2)
-K^\mu(\varepsilon_1\cdot\varepsilon_2).
\label{eq:HDYM_AB_vertex}
\end{align}
where $A_k$ denotes the kth massless gluon and $B_j$ denotes the jth Merlin particle. The second term is longitudinal with respect to the off-shell leg.  Projecting onto the
physical transverse part of the off-shell current yields
\begin{align}
J_{12,\perp}^\mu(A_1,B_2)
&=
\frac{-K^2/M^2}{K^2(1-K^2/M^2)}
P_T^\mu{}_{\nu}(K)
\mathbb V_{\rm YM}^\nu(A_1,B_2)
\nonumber\\
&=
\frac{P_T^\mu{}_{\nu}(K)
\mathbb V_{\rm YM}^\nu(A_1,B_2)}{K^2-M^2}.
\label{eq:HDYM_AB_current}
\end{align}
Thus the physical transverse current has no massless pole but does have a
massive pole,
\begin{equation}
\operatorname{Res}_{K^2=0}J_{12,\perp}(AB)=0,
\qquad
\operatorname{Res}_{K^2=M^2}J_{12,\perp}(AB)\neq0.
\label{eq:HDYM_AB_residues}
\end{equation}
Thus, this gives us the simplest 2-Merlin one gluon amplitude which is non zero because the underlying BG current is non-zero.
If the off-shell leg is projected onto an ordinary massless gluon, the residue
vanishes and one recovers a one-Merlin amplitude with zero value.  If the same
current is projected onto the massive pole, the residue is nonzero and gives an
amplitude with two external Merlins.  This current is therefore genuinely
different from the all-massless BG current, even though both arise from the
same higher-derivative field. At higher rank one may similarly study currents with one on-shell Merlin and an
arbitrary number of on-shell massless gluons.  Their massless-off-shell residues are
constrained by the one-Merlin selection rule, while their massive-off-shell leg residues
probe amplitudes with two external Merlins.  The rank-two calculation above is
the simplest explicit realization of this pole separation.

\subsection{Two on-shell Merlins and the first mixed-branch current}
\label{subsec:HDYM_twoMerlin_current}

We next take both on-shell legs to lie on the massive pole,
\begin{equation}
k_1^2=k_2^2=M^2,
\qquad
k_i\cdot\varepsilon_i=0.
\end{equation}
For $s_{12}=(k_1+k_2)^2$, the longitudinal term in
Eq.~\eqref{eq:HDYM_cubic_map} vanishes because $P^2=Q^2=M^2$, while the
coefficient of the Yang--Mills tensor becomes
\begin{equation}
1-\frac{s_{12}+2M^2}{M^2}
=-\left(1+\frac{s_{12}}{M^2}\right).
\end{equation}
Hence
\begin{equation}
\mathbb V_{\rm HD}^\mu(B_1,B_2)
=
-\left(1+\frac{s_{12}}{M^2}\right)
\mathbb V_{12}^\mu,
\qquad
\mathbb V_{12}^\mu
\equiv
\mathbb V_{\rm YM}^\mu(k_1,k_2;\varepsilon_1,\varepsilon_2).
\end{equation}
The complete higher-derivative current is
\begin{align}
\widehat J_{12}^\mu(B_1,B_2)
&=
-\frac{1+s_{12}/M^2}
{s_{12}(1-s_{12}/M^2)}
\mathbb V_{12}^\mu
\nonumber\\
&=
-\frac{\mathbb V_{12}^\mu}{s_{12}}
+\frac{2\mathbb V_{12}^\mu}{s_{12}-M^2}.
\label{eq:HDYM_BB_current}
\end{align}
This is the first current in which both the massless and massive poles survive
simultaneously.

\subsection{Three- and four-Merlin amplitudes as checks of the BG currents}
\label{subsec:HDYM_Merlin_amplitudes}
The Merlin amplitudes are discussed in details in \cite{Menezes}. Here, we cross check that the amplitudes we obtain from the BG currents match with the ones in that reference paper.
\subsubsection{Three-Merlin amplitude}

The Merlin component of \eqref{eq:HDYM_BB_current} is
\begin{equation}
B_{12}^\mu[BB]
=\frac{2\mathbb V_{12}^\mu}{s_{12}-M^2}.
\end{equation}
Amputating the root at the massive pole multiplies by the inverse residue
$-(s_{12}-M^2)$.  Therefore
\begin{align}
\mathcal A_3(B_1,B_2,B_3)
&=
\lim_{s_{12}\to M^2}
\left[-(s_{12}-M^2)\right]
\varepsilon_{3\mu}B_{12}^\mu
\nonumber\\
&=
-2g\,\varepsilon_{3\mu}\mathbb V_{12}^\mu.
\label{eq:HDYM_A3_BBB}
\end{align}
Thus the three-Merlin amplitude has the ordinary Yang--Mills cubic Lorentz
structure but carries the characteristic relative coefficient $-2$.

\subsubsection{Three-particle Merlin current}

For three external Merlins and a Merlin off-shell leg, the pair $(12)$ may first produce
either the massless current which is the massless branch part of the last line of \eqref{eq:HDYM_BB_current} or the massive current, which is the second term of the last line of the same equation.
The same occurs in the $(23)$ channel. Let us define
\begin{equation}
\Pi_{123}^{\mu\nu}
=
\eta^{\mu\nu}
-\frac{K_{123}^\mu K_{123}^\nu}{M^2},
\end{equation}
and
\begin{align}
\mathcal X_{12|3}^{\nu}
&\equiv
\mathbb V_{\rm YM}^{\nu}
(K_{12},k_3;\mathbb V_{12},\varepsilon_3),
\\
\mathcal X_{1|23}^{\nu}
&\equiv
\mathbb V_{\rm YM}^{\nu}
(k_1,K_{23};\varepsilon_1,\mathbb V_{23}).
\end{align}
Using the two-point currents in \eqref{eq:HDYM_BB_current}, one finds
\begin{align}
B_{123}^{\mu}
=
\frac{\Pi_{123}^{\mu}{}_{\nu}}{s_{123}-M^2}
\Bigg[{}
&\left(-\frac1{s_{12}}+\frac4{s_{12}-M^2}\right)
\mathcal X_{12|3}^{\nu}
\nonumber\\
&+\left(-\frac1{s_{23}}+\frac4{s_{23}-M^2}\right)
\mathcal X_{1|23}^{\nu}
+3\mathbb V_{4,{\rm YM}}^{\nu}(1,2,3)
\Bigg].
\label{eq:HDYM_B123}
\end{align}
The current is manifestly different from an ordinary Yang-Mills BG current.
It contains both $1/s_{ij}$ and $1/(s_{ij}-M^2)$ exchange poles.  The factor
$4$ multiplying the massive pole arises from the product of the two relative
$BBB$ vertex factors, together with the sign of the massive
propagator.

\subsubsection{Four-Merlin amplitude}

Putting the off-shell leg on the massive shell,
\begin{equation}
K_{123}=-k_4,
\qquad
k_4^2=M^2,
\qquad
k_4\cdot\varepsilon_4=0,
\end{equation}
and amputating the final massive propagator gives, for the ordering
$(1,2,3,4)$,
\begin{align}
\mathcal A_4(B_1,B_2,B_3,B_4)
={}&
g^2\mathcal R_s
\left(\frac1s-\frac4{s-M^2}\right)
+g^2\mathcal R_t
\left(\frac1t-\frac4{t-M^2}\right)
\nonumber\\
&-3g^2\mathcal A_{4,{\rm ct}}^{\rm YM},
\label{eq:HDYM_A4_BBBB}
\end{align}
where
\begin{equation}
s=s_{12},
\qquad
t=s_{23},
\end{equation}
and
\begin{equation}
\mathcal R_s
=\varepsilon_{4\mu}\mathcal X_{12|3}^\mu,
\qquad
\mathcal R_t
=\varepsilon_{4\mu}\mathcal X_{1|23}^\mu,
\qquad
\mathcal A_{4,{\rm ct}}^{\rm YM}
=\varepsilon_{4\mu}\mathbb V_{4,{\rm YM}}^\mu(1,2,3).
\end{equation}
The characteristic exchange factor
\begin{equation}
\frac1s-\frac4{s-M^2}
\label{eq:HDYM_massive_pole_combination}
\end{equation}
has an immediate BG interpretation: the $1/s$ contribution comes from the
massless component of the $BB$ current, while the $1/(s-M^2)$ contribution
comes from its massive component.  The four-Merlin amplitude therefore serves
as a direct check that the BG current correctly resolves the two propagating
branches of the higher-derivative theory.

\section{Conclusion}
\label{sec:conclusion}

In this work we have developed a Berends-Giele description of higher-derivative
gauge theories, focusing on higher-derivative scalar electrodynamics and
higher-derivative Yang--Mills theory.  Our main aim has been to understand how
the additional poles and interaction structures generated by higher derivatives
are encoded directly at the level of multiparticle off-shell currents and how
the resulting recursion relations distinguish the ordinary massless sector from
the additional massive degrees of freedom.

We first studied higher-derivative scalar QED.  Starting from the
higher-derivative equations of motion, we constructed recursive currents for the
scalar and gauge sectors and compared them with the corresponding expressions
obtained directly from Feynman diagrams.  The agreement between the two
constructions provides a useful consistency check on the perturbiner
formulation.  In particular, the higher-derivative scalar kinetic operator
modifies both the propagator and the interaction vertices, while the photon
sector develops the characteristic higher-derivative pole structure.  The
recursive formulation organizes these modifications naturally in terms of
off-shell currents and makes the relevant Ward identities transparent.

We then turned to higher-derivative Yang--Mills theory with Lagrangian
containing the operator $(D_\mu F^{\mu\nu})^2$.  The gauge-field propagator
contains both the ordinary massless gluon pole and an additional massive pole,
which according to the literature is interpreted as a Merlin mode.  We
derived the corresponding Berends--Giele recursion relations and analyzed the
different branches generated by these two propagating sectors.

A particularly simple structure emerges when all external states are ordinary
massless gluons.  At rank two the higher-derivative factor carried by the cubic
vertex cancels the massive factor in the propagator, leaving precisely the
ordinary Yang--Mills current.  At rank three the cancellation is more involved:
the two exchange diagrams contain explicit higher-derivative corrections, and
the off-shell root initially carries the complete higher-derivative propagator.
After including the higher-derivative quartic contact interaction, however, the
full numerator develops an overall factor $1-s_P/M^2$, which cancels the
massive pole of the propagator of the off-shell leg.  The resulting current is again identical
to the Yang--Mills Berends--Giele current.

We further showed that this property is not restricted to low multiplicity.
Writing the higher-derivative action as the Yang--Mills action supplemented by
a term quadratic in the Yang--Mills equations of motion allows the perturbiner
equations to be organized recursively.  Assuming that all lower-rank massless
currents satisfy the Yang--Mills equations, the higher-derivative corrections
at the next rank reduce to a factor multiplying the same equation of motion.
Away from the massive pole this implies that the current again satisfies the
ordinary Yang--Mills recursion.  Consequently,
\begin{equation}
J_{P}^{\mu,{\rm HD}}
=
J_{P}^{\mu,{\rm YM}}
\end{equation}
for an arbitrary multiparticle word $P$ constructed entirely from massless
on-shell gluon states.
The situation changes once external Merlin states are introduced.  The
multiparticle currents then retain genuine higher-derivative information and
naturally decompose into massless and massive propagator channels.  For a
current containing one massless gluon and one Merlin state, the massless pole
drops out and the resulting current propagates only through the massive branch.
This immediately explains why a purely gluonic amplitude containing exactly
one external Merlin vanishes when the final state is extracted as the residue
at the massive pole.  In contrast, currents with two external Merlins contain
both massless and massive poles, providing the elementary branching structure
from which higher-Merlin amplitudes are generated.

As further checks of the recursion, we extracted the three- and four-Merlin
tree amplitudes from the corresponding BG currents.  The three-point result
reproduces the characteristic relative coefficient of the cubic Merlin
interaction, while the four-point amplitude contains both massless and massive
exchange poles with the expected relative weight,
together with the modified quartic contact interaction.  These expressions
agree with the previously known higher-derivative Yang--Mills amplitudes and
therefore provide a nontrivial consistency check on the current-level construction.\\~\\
Our construction maybe helpful for several future research directions. First, off-shell amplitudes like the BG currents are instrumental to develop loop level integrands via sewing two off-shell legs of the current and using the recursive method. The relation that we found regarding the equality of the YM and HDYM BG currents can help us understand the relations of 1-loop integrands of YM and HDYM theories. The second point is about the colour/kinematics duality. In \cite{Fu}, it has been shown that one can construct the so called YM bracket to understand the duality at four points, and interpret it as an algebra of the Drinfeld double. Such a bracket construction can also been done for HDYM, which may help us why the duality works in this case. Further, this can be investigated at the one loop integrand level and help us construct loop level integrands in quadratic gravity from HDYM. We leave this to a future publication.
\section{Acknowledgements}
We thank Yi Xiao Tao for multiple discussions about the project. During the preparation of the manuscript, the author used the Open AI chatgpt software engine
to check symbolic computations, text formatting and references list completeness. After using
this tool, the author reviewed and edited the content as needed and bears full responsibility for the content of the
published article.

\section{Appendix}
We sketch the relevant terms of the Lagrangian in \eqref{eq:HDLag} to derive the higher order scalar-photon vertices. We define
\begin{equation}
\mathcal{Q}\equiv-\Dhat^2
=\mathcal{Q}_0+g\mathcal{X}+g^2\mathcal{Y},
\qquad
\mathcal{Q}_0=-\partial^2,
\end{equation}
where
\begin{equation}
\mathcal{X}
=\ii\left(2\Ahat\cdot\partial+\partial\cdot\Ahat\right),
\qquad
\mathcal{Y}=\Ahat^2.
\label{eq:XYdef}
\end{equation}
Then
\begin{align}
\Kphi(\mathcal{Q})
={}&\Kphi(\mathcal{Q}_0)
+g\mathcal{K}_1
+g^2\mathcal{K}_2
+g^3\mathcal{K}_3
+g^4\mathcal{K}_4,
\nonumber
\\[1mm]
\mathcal{K}_1
={}&\mathcal{X}
-\frac1{M_\phi^2}
(\mathcal{Q}_0\mathcal{X}+\mathcal{X}\mathcal{Q}_0),
\nonumber
\\
\mathcal{K}_2
={}&\mathcal{Y}
-\frac1{M_\phi^2}
(\mathcal{Q}_0\mathcal{Y}+\mathcal{Y}\mathcal{Q}_0+\mathcal{X}^2),
\nonumber
\\
\mathcal{K}_3
={}&-\frac1{M_\phi^2}
(\mathcal{X}\mathcal{Y}+\mathcal{Y}\mathcal{X}),
\nonumber
\\
\mathcal{K}_4
={}&-\frac1{M_\phi^2}\mathcal{Y}^2.
\label{higher vertices}
\end{align}
Using these, the cubic and quartic Feynman rules can be extracted rather easily. We do not write the explicit forms as it is not neccessary for our purposes.

\end{document}